\documentclass[12pt]{revtex4}
\usepackage{graphicx}
\usepackage{rotating}
\usepackage[english]{babel}
\usepackage{times}
\usepackage{mathptm}
\usepackage{supertabular}
\usepackage{dcolumn}
\usepackage{fancyhdr}

\begin{document}

\begin{titlepage}
\title{  B(E2)$\uparrow$ PREDICTIONS FOR  EVEN-EVEN NUCLEI IN THE 
DIFFERENTIAL EQUATION MODEL$*$} 

\author{{R. C. Nayak} \\
{Department of Physics, Berhampur University, Berhampur-760 007, India} \\
{ S. Pattnaik} \\
{ Taratarini  College, Purusottampur, Ganjam, Odisha, India.}}

\begin{abstract}
  We use the recently developed Differential Equation  Model  for the
 reduced electric quadrupole 
transition probability
 B(E2)$\uparrow$ for predicting its  values for a wide range of  even-even 
nuclides
almost  throughout the nuclear landscape from Neon to Californium.
 This is made  possible  as the principal equation in the model, namely,
the  differential equation connecting the B(E2)$\uparrow$ value of a given
nucleus 
  with its 
derivatives with respect to neutron and proton numbers 	 
provides two different recursion relations, each
connecting   three different  neighboring even-even 
nuclei from lower to
 higher mass 
numbers and vice-verse. These relations  helped us to extrapolate 
 from known to unknown terrain of
the B(E2)$\uparrow$ landscape and thereby facilitate its predictions
throughout.

\end{abstract}
\maketitle
\vskip 5cm
$*$ This is a slightly modified version of the article submitted for
 publication  in Atom. Data and Nucl. Data Tabl.(2014).

\end{titlepage}

\newpage
\section{  INTRODUCTION}

Reduced electric quadrupole transition probability B(E2)$\uparrow$   plays 
an important role for the study of nuclear structure, especially the deformation
properties of nuclei. 
Such studies got a  boost with the  
 advent of isotope facilities providing a large amount of  experimental 
data for several nuclides throughout the nuclear chart.
 The existence of a large  volume  of experimental data  lead Raman et al.
 \cite{rmn} at the Oak Ridge 
Nuclear Data Project \cite{rmn,rm3} to make a comprehensive
analysis of all those data in  preparing  the most sought-after experimentally
 adopted data table.
Of late,  
Pritychenko et al. \cite{prt} followed the process in compiling the newly emerging  data
sets  for some even-even
nuclei  near  $ N \sim Z \sim $ 28. These new data including the old set obviously put a 
challenge for the nuclear theorists to understand them.

Theoretically, possible existence of nuclear symmetry in the properties of 
nuclei led Ross and Bhaduri\cite{rbh} in developing difference
equations involving the B(E2)$\uparrow$ values of neighboring even-even nuclei.
Patnaik et al. \cite{pat}
 on the other hand have  also succeeded   in establishing
even more  simpler difference equations connecting  these values of
four neighboring even-even nuclei. Apart from these difference equations, there
exists several nuclear   models such as Single-Shell Asymptotic Nilson Model 
(SSANM) \cite{ssa} and  Finite-Range Droplet Model (FRDM)
\cite{frd}
in predicting B(E2)$\uparrow$ values to some success. 
 
 Just recently,  we have succeeded in  developing \cite{dem} a new model for
 B(E2)$\uparrow$
termed as the Differential Equation Model (DEM) according to which,
the B(E2)$\uparrow$  value of 
 a given nucleus is expressed in terms of  its derivatives 
with respect to the neutron and proton numbers.
   In fact, we had already demonstrated \cite{frd} its 
 goodness in reproducing the known \cite{rmn} B(E2)$\uparrow$ values throughout and 
also in predicting the unknown data of Pritychenko et al. \cite{prt} with 
success. It is worth mentioning here,  
 that we could visualize such a differential equation on the basis of a similar
 equation  
for  an important component of the the ground-state energy of a nucleus,
termed as the local energy
 $\eta$ of the recently developed   Infinite Nuclear Matter
(INM) model \cite{inm,in2,in3,in5,rcn} of atomic nuclei.
More over,  the $\eta$-relation connecting the partial
 derivatives of  the local energies in the INM model 
   has been shown to be primarily responsible for the success \cite{in5} of
 the INM 
model as a mass-formula. 
The philosophy  that  
 any relation in the form of a differential equation 
is  sound enough to possess a good predictive ability was
  well demonstrated in the INM model,  specifically for
the prediction \cite{in5} of nuclear masses 
throughout the nuclear chart.    
Since the local energy  basically constitutes  the
 shell and 
deformation energies,  and has been shown \cite{rcn} to carry
 also the shell
structure of a given nucleus, 
   is  expected to have  
  a good one-to-one correspondence  with the properties of 
excited states,  and in particular 
 the reduced transition probabilities. 
All these aspects therefore lead us in formulating the
  Differential Equation Model \cite{dem} for B(E2)$\uparrow$.

In the following Section 2,  we briefly discuss the model and its salient 
features. Section 3 covers the scheme of  B(E2)$\uparrow$ predictions
throughout nuclear landscape 
 and   calculation of the associated
uncertainties
in them. In Section 4, we  present all our DEM   predictions
along with
the known experimentally adopted B(E2)$\uparrow$ data.

\section{ THE DIFFERENTIAL EQUATION MODEL  FOR B(E2)$\uparrow$}

The Differential Equation Model  for B(E2)$\uparrow$  has been well 
	described elsewhere\cite{dem}. Here we just highlight its 
salient features 
 for sake of completeness and also to facilitate  discussion
of our  predictions. The principal equation in the model  termed as the 
DEM equation for B(E2)$\uparrow$  is given by

\begin{equation}
  \label{be2} {B(E2)[N,Z]/ A}={1\over 2} \left[(1+\beta){\Bigl(\partial B(E2)
  / \partial N\Bigr)}_Z + (1-\beta){\Bigl(\partial B(E2)/  \partial Z}\Bigr)_N\right],
\end{equation}
where Z, N and  A are the usual proton, neutron and mass numbers of a given 
nucleus.
$\beta$ as usual is the symmetry parameter (N-Z)/(N+Z).
This DEM Eq. (\ref{be2}) as we see,  connects the B(E2)$\uparrow$
 value of a given nucleus (N,Z)
with its partial derivatives with respect to neutron   and proton 
numbers N and  Z. 
In order to utilize the differential  Eq. (\ref{be2}) for all practical
 purposes, we have used  the usual forward and backward definitions of the 
two derivatives occurring in the equation  as
\begin{eqnarray}
  \label{der} 
\Bigl({\partial B(E2)/
  \partial N}\Bigr)_Z &\simeq&{1\over 2}  \Bigl[ B(E2)[N+2,Z]-B(E2)
  [N,Z]\Bigr], \nonumber \\ 
\Bigl( {\partial
  B(E2)/ \partial Z}\Bigr)_N &\simeq&{1\over 2}  \Bigl[ B(E2)
  [N,Z+2]-B(E2)[N,Z] \Bigr] ,\\
& & and \nonumber \\
\Bigl({\partial B(E2)/
  \partial N}\Bigr)_Z &\simeq&{1\over 2}  \Bigl[ B(E2)[N,Z]-B(E2)
  [N-2,Z]\Bigr], \nonumber \\ 
\Bigl( {\partial
  B(E2)/ \partial Z}\Bigr)_N &\simeq&{1\over 2}  \Bigl[ B(E2)
  [N,Z]-B(E2)[N,Z-2] \Bigr] .
\end{eqnarray}
Substitution of these definitions in the DEM equation (\ref{be2}),
 led us to obtain 
two   different recursion relations in  B(E2)$\uparrow$
connecting three neighboring  even-even nuclei in (N,Z) space. These are
\begin{eqnarray}
\label{b2f}
B(E2)[N,Z] &=& {N \over {A-2}}\: B(E2) [N-2,Z] + {Z \over {A-2}}\:B(E2) 
[N,Z-2]  ,\\
\label{b2b}
 B(E2) [N,Z] &=& {N \over {A+2}}\: B(E2) [N+2,Z]+{Z \over {A+2}}\: B(E2)
[N,Z+2] .
\end {eqnarray}
We see that the first recursion  relation (\ref{b2f}) connects B(E2)$\uparrow$
values of the nuclei (N,Z), (N-2,Z)
and (N,Z-2), while the second one (\ref{b2b}) connects those of (N,Z), 
(N,Z+2) and (N+2,Z). 
  It is also  obvious to see that    the  first one  relates  B(E2)$\uparrow$ values  
of lower  to higher mass nuclei while the second one 
relates higher to lower mass.
Thus   depending on the availability of $B(E2)\uparrow$ data, one can use 
either or both  of these two relations to obtain the corresponding
unknown  values of   neighboring nuclei. 

The validity of these equations at numerical level has been already
 demonstrated\cite{dem}
with the use of the experimental data set of Raman et al.\cite{rmn}. More 
importantly,
 the goodness of the model predictions has been also shown\cite{dem} 
   when compared  with the latest experimental  data set
 of Pritychenko et al.\cite{prt} for  all the isotopes of Cr, Fe, Ni and Zn.

\section{ SCHEME OF   B(E2)$\uparrow$ PREDICTIONS IN THE MODEL }

Each of the two recursions relations 
 (\ref{b2f}) and (\ref{b2b}) derived from the DEM  Eq. (\ref{be2}) connects
 B(E2)$\uparrow$ values of three different neighboring nuclei as stated earlier.
More importantly,  each of these
relations can be expressed  in three different ways just by interchanging
 the three
terms occurring in  them from left to right and vice-verse. 
Thus altogether,  these two relations in principle can act as six recursion
relations as far as a particular nucleus is  concerned,  and consequently 
can generate  up to  six alternate
values  for the    nucleus. Of course the exact number of alternate 
values for a given nucleus  depends on
the availability of  the  neighboring input data.
Since each of these values is equally probable, 
the predicted value for a given nucleus is  then obtained by the arithmetic  
mean of all those  generated values so obtained. 

Thus, following the procedure as laid down above, our scheme in the first step
generates  all
 possible  B(E2)$\uparrow$ values of both known as well as the  unknown
neighboring even-even nuclides using the experimentally 
available data. These predictions so obtained in the first step may be termed as
 the first generation predictions.
Then in the second step, these  predictions  along with 
 the  experimental  data set are again used to obtain the  second generation 
 predictions. 
  In principle, this scheme can be continued for all possible extrapolations.
  However in practice, this scheme cannot be continued for more 
number
of extrapolations  due to  accumulation of errors in the 
predicted  values. 
Accordingly one has to compromise in between the amount of error accumulation 
and the order  of extrapolations.

It is  desirable  to  calculate possible uncertainties or errors  associated 
with  the predictions of our model.
 Since the  model basically  uses the recursion relations involving the data
points occurring in the right hand sides of the  relations, the uncertainties
 associated with  the predictions can be easily evaluated taking into account
 those of 
the data points  following the usual error analysis procedure.
Therefore to start with  the  predictions of  the first generation,
 the experimental uncertainties of the input data points are taken into account
in evaluating  the corresponding associated errors.
  The same methodology
is again followed for the subsequent extrapolations, taking into account
  the evaluated 
uncertainties of the previously  generated DEM predictions along with those of
 the experimental data
involved in  the predictions. It is needless to say that errors associated with
the predictions would
obviously increase from generation to generation.

\section{ RESULTS AND  DISCUSSIONS }

Following the scheme outlined in the previous section, we have evaluated
 the DEM predictions  using the combined
 data set 
 of Raman et al. \cite{rmn} and  Pritychenko et al. \cite{prt}. 
It may be noted that Pritychenko et 
al. have recently compiled the adopted  values for the isotopes of
Cr, Fe, Ni and Zn using the latest experimental data. Therefore the old data set of Raman et al. \cite{rmn} for these 
isotopes has been replaced by 
 those of  Pritychenko et al. \cite{prt}. As a result,
 the combined data set comprises altogether 325 even-even nuclides spread over
the  nuclear landscape ranging from Ne to Cf (
 Z=10 to 98).
Our DEM predictions are also confined to all the
possible even-even isotopes of the same isotope series 
with extrapolations up to almost 3 to 5 steps on either
 side  of the known data base.  Thus altogether, we have  predicted  the   
B(E2)$\uparrow$ values  of 
  another 251 nuclides going up to 3rd order  of extrapolations. The 
DEM predictions along with the combined experimental data set
 where  available,  are presented in Table 1.  
 For sake of analyzing the degree of agreement with experiment,
we  followed Raman et al. \cite{rmn}
in presenting  our analysis in terms of  the ratio of DEM values with those  of 
the experiment in Fig. 1. As we can see from this figure,  the 
	percentage of agreement is 93\%  which is quite remarkable, and may
be termed excellent as per the yardstick formulated by Raman et al. \cite{rmn}.
Accordingly, such excellent agreement of the DEM predictions with experiment obviously
commands  confidence in our other predictions given in Table 1.  

\begin{figure}[h!t!b!]
\includegraphics[width=5in,height=4in,angle=0]{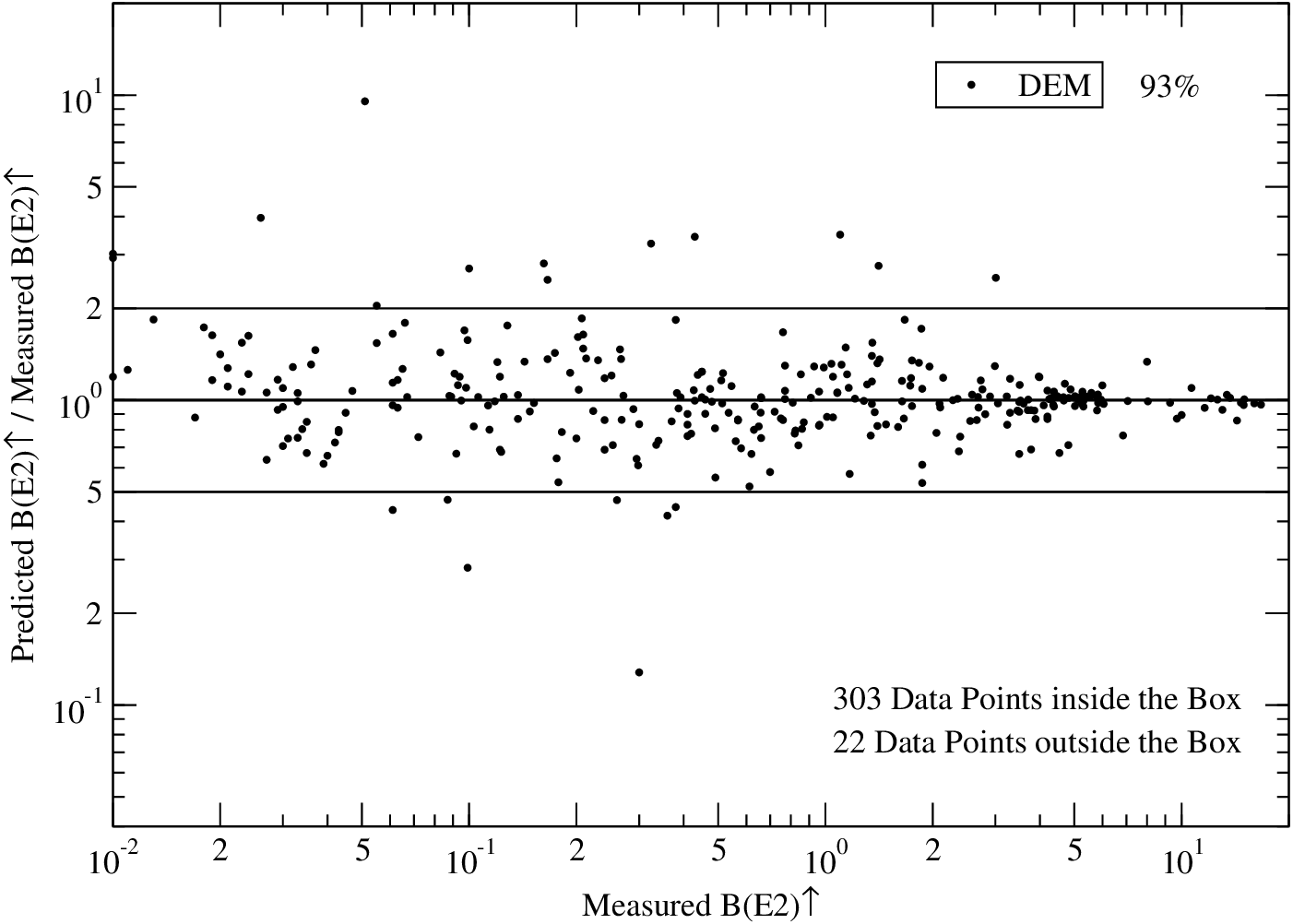}
\caption{Comparison between the measured B(E2)$\uparrow$ values and those
of the DEM predictions. The values inside the box agree within a factor of two.
The percentage of values lying within the region of agreement is shown on
the graph. }
\end{figure}

In order to provide a better  perception of the DEM predictions in contrast
to those of experiment, we also present them 
graphically in Figs. 2 to 7
for almost all those isotope series showing significant changes in nuclear structure.
 Also
included in these figures, predictions of two other  standard theoretical
models namely  
SSANM \cite{ssa} and FRDM \cite{frd} for sake of good comparison. 
First of all one can  see from these figures,  the excellent agreement of our DEM predictions with those
of the experimental data. Secondly,  in most of the cases we see, that the DEM 
predictions almost follow the experimental trends. 
Even the sharply changing nuclear structures in most of the cases are well 
reproduced by our DEM predictions compared to the other two theoretical models
SSANM \cite{ssa} and FRDM \cite{frd}.

{99}

\clearpage

\section{EXPLANATION OF TABLE}

\begin{center}
\begin{tabular}{ll}
$Z$ & Proton number. The $B(E2)\uparrow$ data table  is ordered by increasing proton number.\\
        & The corresponding name of each element is given in parenthesis. \\

$N$	& Neutron number. \\

$A$	& Mass number. \\

B(E2)$\uparrow$ (DEM) & Predicted values of B(E2)$\uparrow$ s  in the Differential
Equation Model (DEM) 
of a nucleus (N,Z) followed by its  model uncertainty in units of $e^2b^2$.\\

B(E2)$\uparrow $ (EXP)& Experimentally adopted values \cite{rmn,prt}
of  B(E2)$\uparrow$ 
of a nucleus (N,Z) followed by its   uncertainty in units of $e^2b^2$.\\
 \\
\end{tabular}
\end{center}

\section{EXPLANATION OF FIGURES 2-7}

{\bf Summary of Graphs of B(E2)$\uparrow$ Predictions for Ne to Cf (Z=10-98)
		Isotopes}

These figures show plots of  all the DEM predictions marked as DEM, and the 
experimentally adopted \cite{rmn,prt} data labeled as EXP where available,
 for
most of the isotope series showing significant nuclear structure. Also included in them, the 
predictions of two standard theoretical models SSANM\cite{ssa} and FRDM 
\cite{frd} labeled as SSA and FRD respectively just for sake of comparison. 

\begin{figure}[ht]
\includegraphics[width=\linewidth,height=22cm]{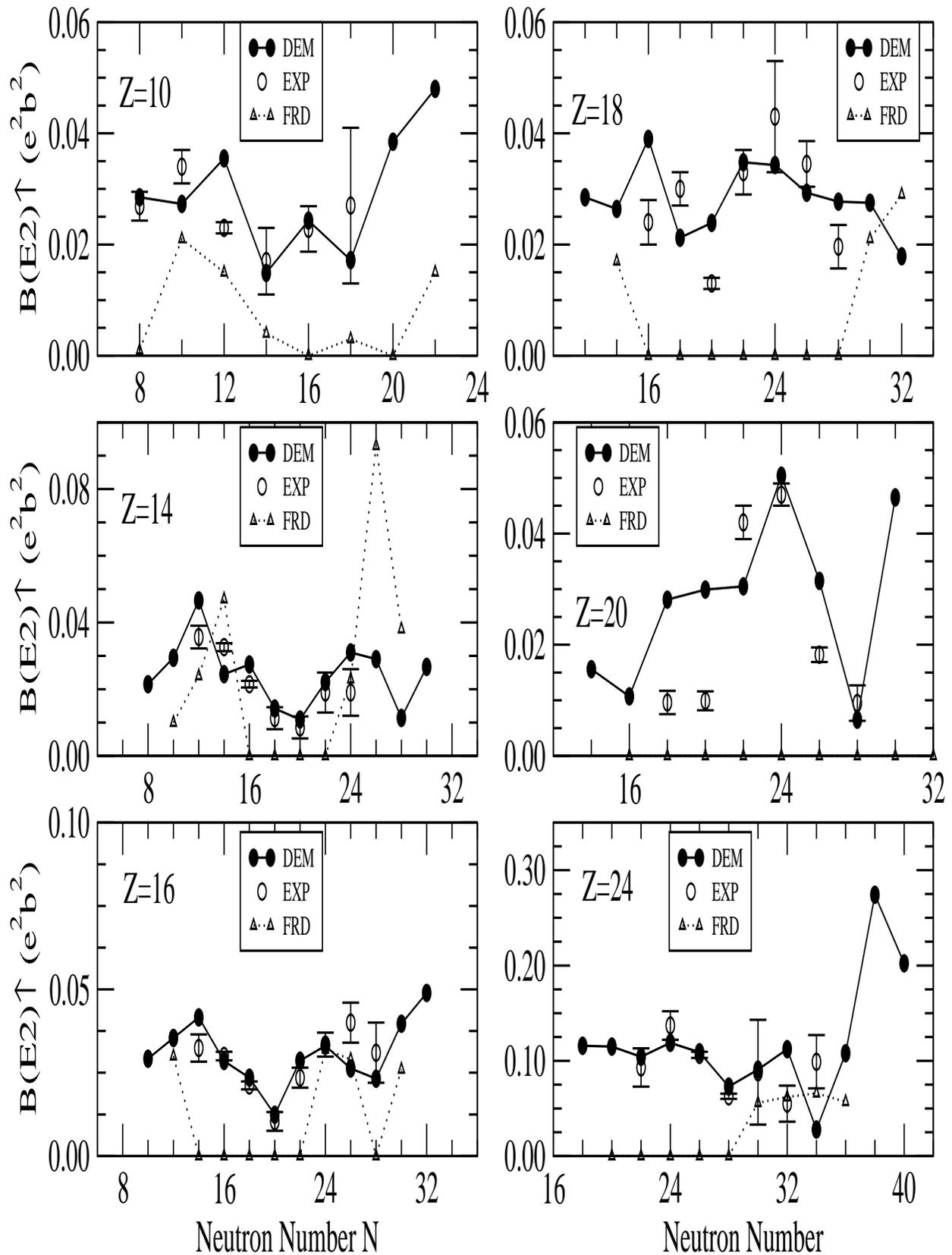}
\caption{Plots of B(E2)$\uparrow$ Predictions marked as DEM compared with 
experiment (EXP) where available. Also shown predictions of SSANM and FRDM
marked as SSA and FRD respectively.}
\end{figure}
\clearpage
\begin{figure}[ht]
\includegraphics[width=\linewidth,height=22cm]{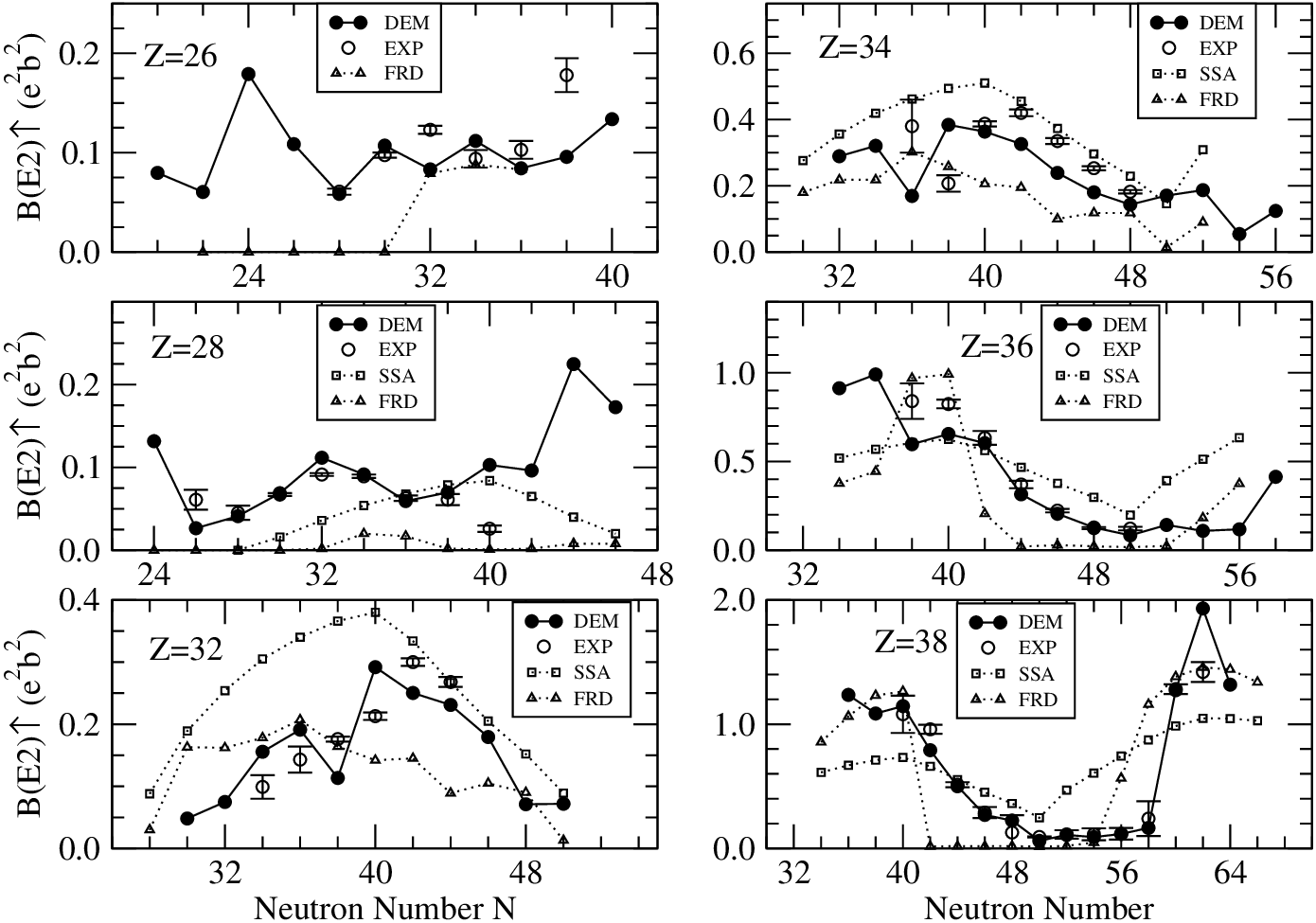}
\caption{Plots of B(E2)$\uparrow$ Predictions marked as DEM compared with 
experiment (EXP) where available. Also shown predictions of SSANM and FRDM
marked as SSA and FRD respectively.}
\end{figure}
\clearpage
\begin{figure}[ht]
\includegraphics[width=\linewidth,height=22cm]{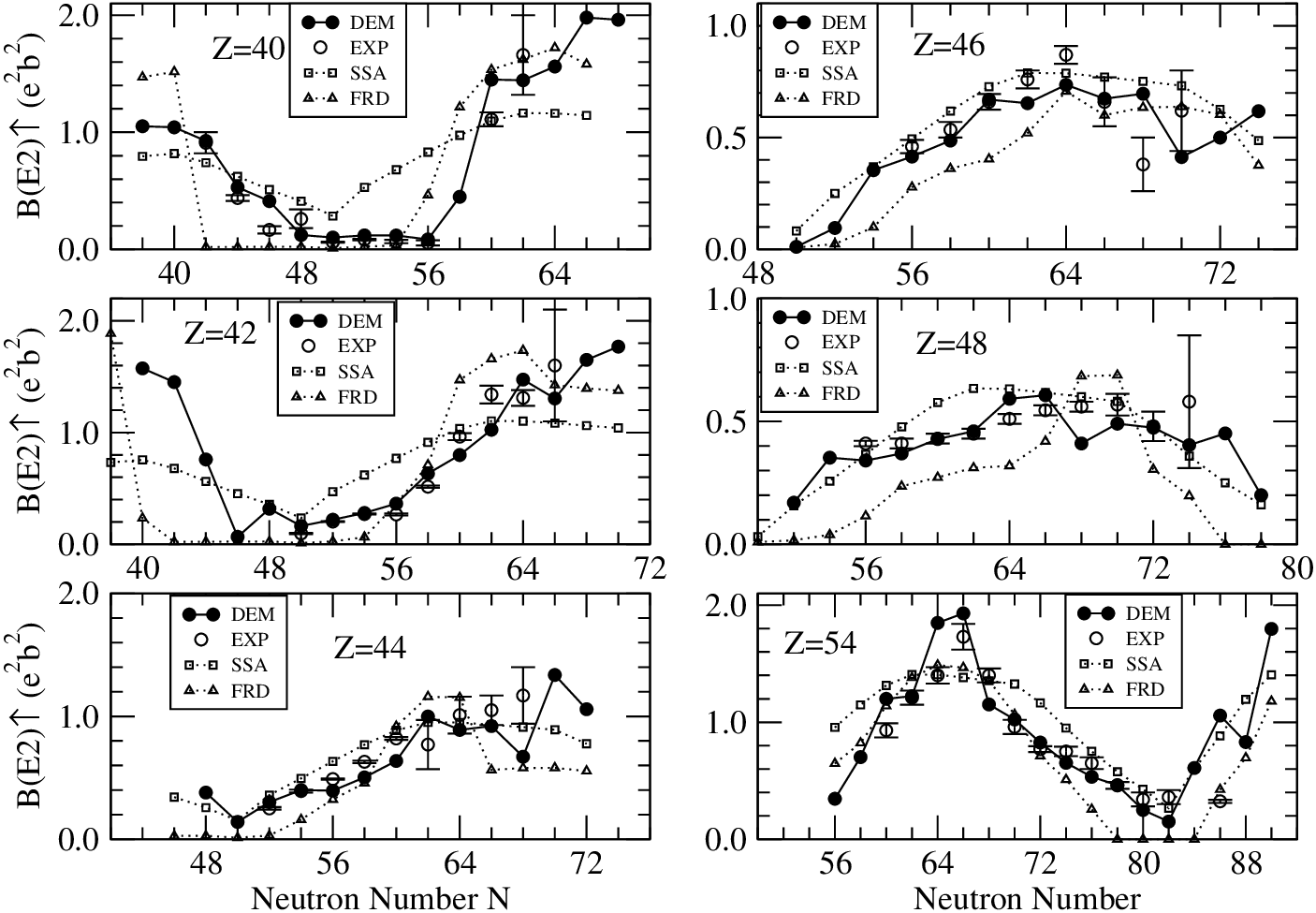}
\caption{Plots of B(E2)$\uparrow$ Predictions marked as DEM compared with 
experiment (EXP) where  available. Also shown predictions of SSANM and FRDM
marked as SSA and FRD respectively.}
\end{figure}
\clearpage
\begin{figure}[ht]
\includegraphics[width=\linewidth,height=22cm]{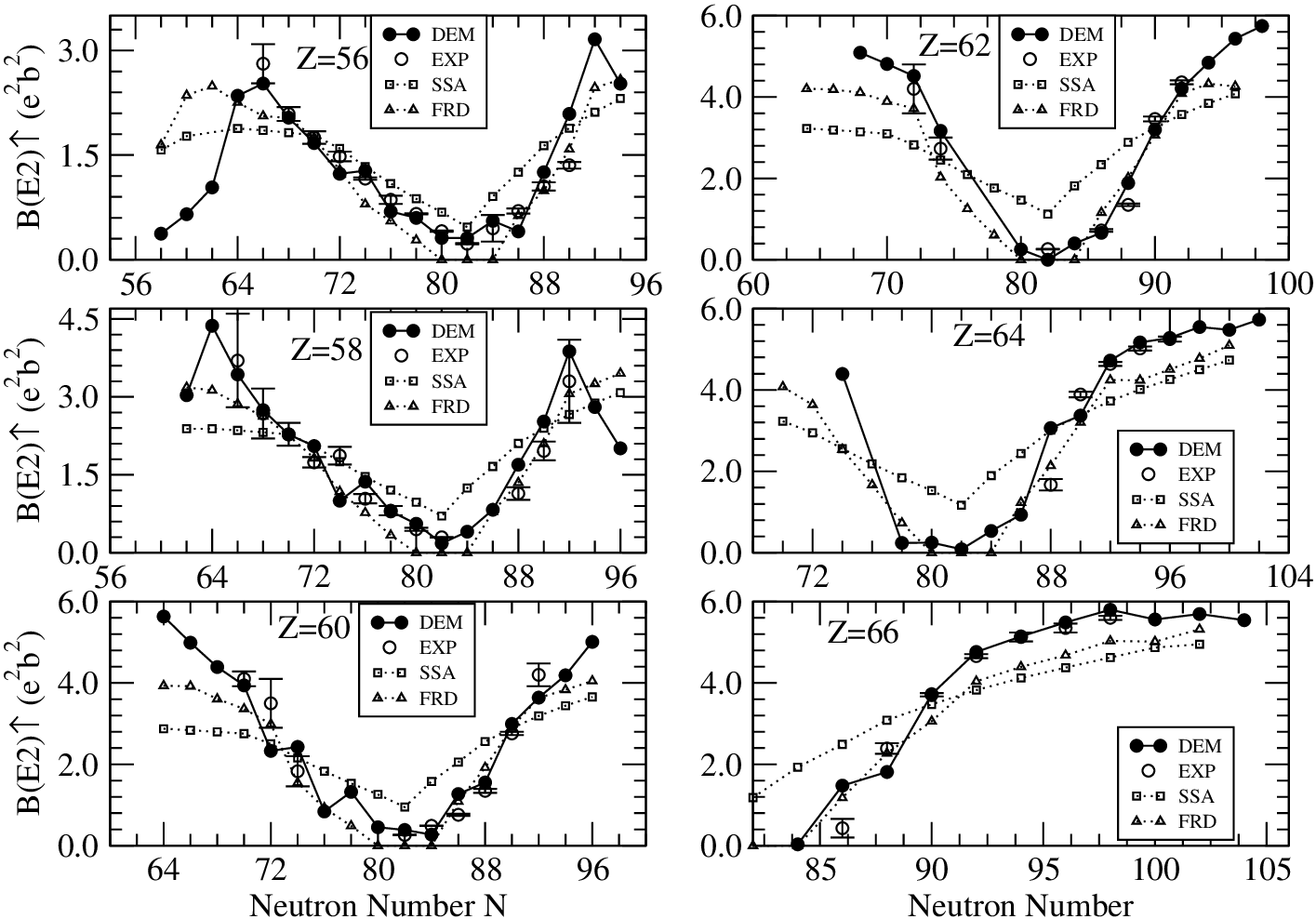}
\caption{Plots of B(E2)$\uparrow$ Predictions marked as DEM compared with 
experiment (EXP) where  available. Also shown predictions of SSANM and FRDM
marked as SSA and FRD respectively.}
\end{figure}
\clearpage

\begin{figure}[ht]
\includegraphics[width=\linewidth,height=22cm]{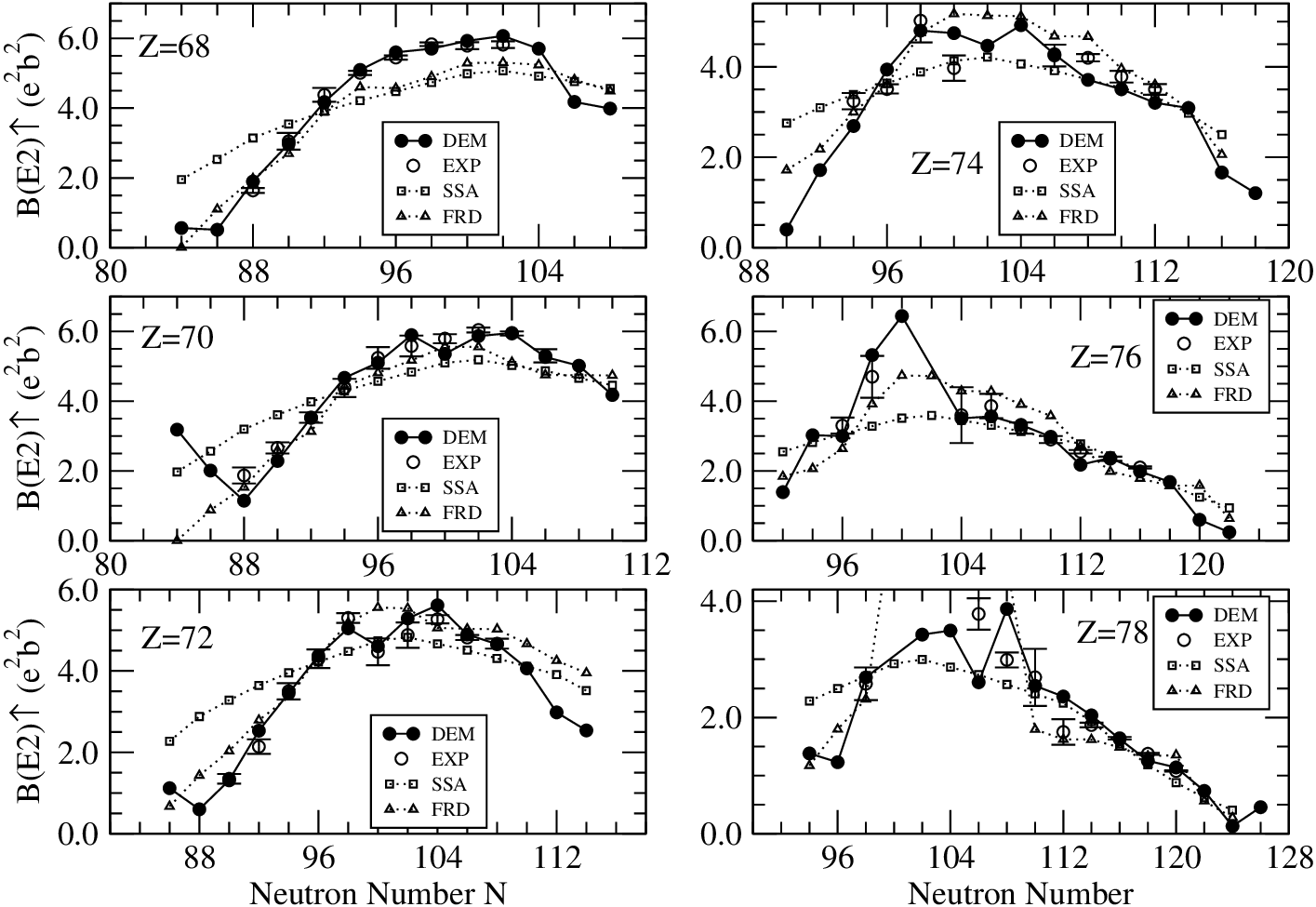}
\caption{Plots of B(E2)$\uparrow$ Predictions marked as DEM compared with 
experiment (EXP) where available. Also shown predictions of SSANM and FRDM
marked as SSA and FRD respectively.}
\end{figure}
\clearpage
\begin{figure}[ht]
\includegraphics[width=\linewidth,height=22cm]{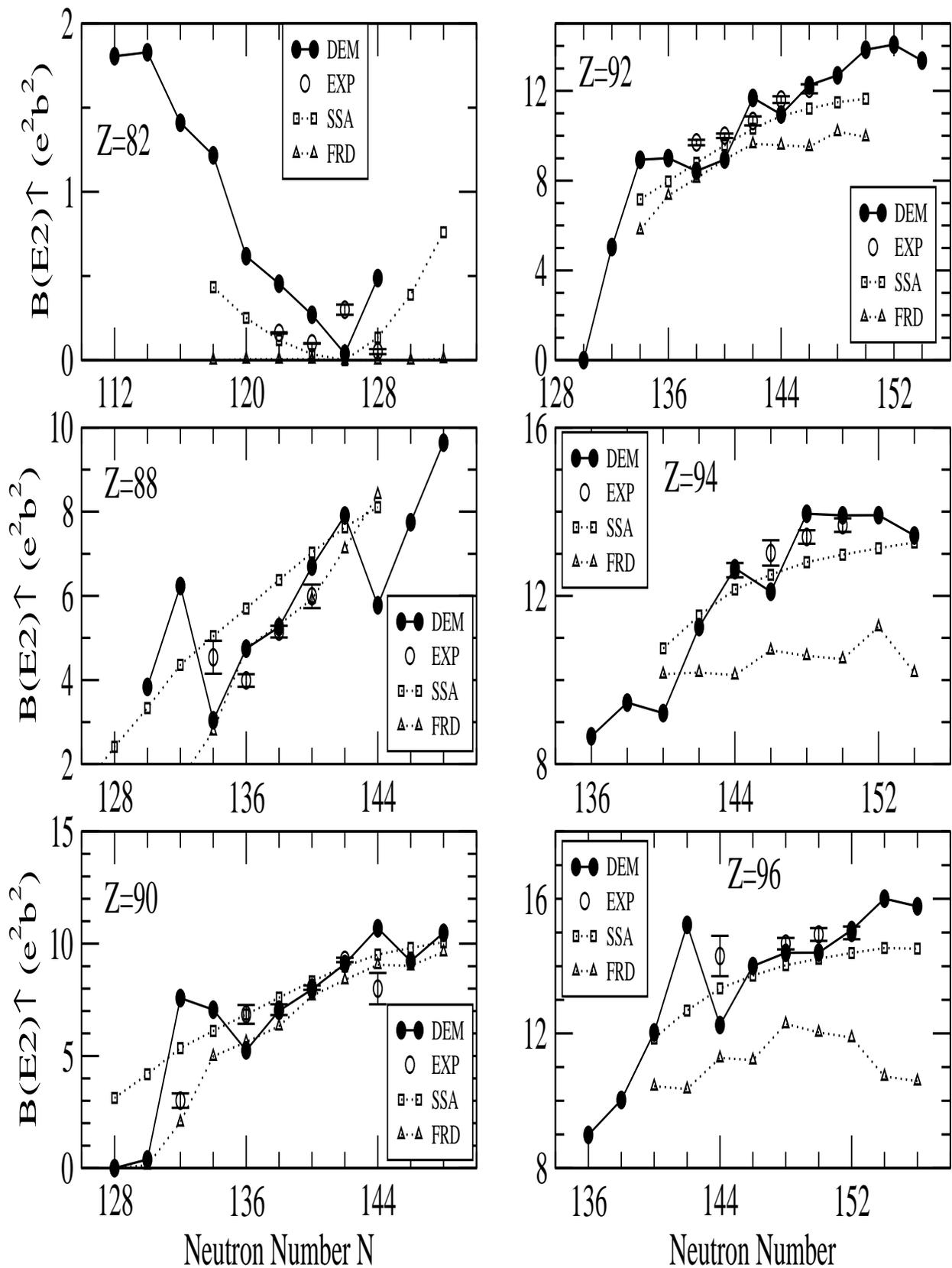}
\caption{Plots of B(E2)$\uparrow$ Predictions marked as DEM compared with 
experiment (EXP) where available. Also shown predictions of SSANM and FRDM
marked as SSA and FRD respectively.}
\end{figure}
\clearpage
\section{ TABLE}
	{\bf {Predicted Values of  $B(E2)\uparrow$ (DEM)
          Compared to  Experiment (EXP) Where Available}} 
\end{document}